\documentclass[reprint,superscriptaddress,amsmath,amssymb,aps,prr,floatfix]{revtex4-2}
\usepackage[usenames, dvipsnames]{color}

\usepackage{graphicx}
\usepackage{dcolumn}
\usepackage{bm}
\usepackage{siunitx}
\usepackage{hyperref}

\usepackage[capitalize]{cleveref}
\hypersetup{
    colorlinks,
    citecolor=blue,    filecolor=blue,
    linkcolor=blue,    urlcolor=blue
}
\usepackage{booktabs}

\begin{document}

\title{Multi-objective and multi-fidelity Bayesian optimization of laser-plasma acceleration}

\author{F. Irshad}
\affiliation{Ludwig-Maximilian-Universit\"at M\"unchen, Am Coulombwall 1, 85748 Garching, Germany}
\author{ S. Karsch}

\author{A. D\"opp}

\affiliation{Ludwig-Maximilian-Universit\"at M\"unchen, Am Coulombwall 1, 85748 Garching, Germany} 
\affiliation{Max Planck Institut für Quantenoptik, Hans-Kopfermann-Strasse 1, Garching 85748, Germany}

\begin{abstract}

Beam parameter optimization in accelerators involves multiple, sometimes competing objectives. Condensing these individual objectives into a single figure of merit unavoidably results in a bias towards particular outcomes, in absence of prior knowledge often in a non-desired way. Finding an optimal objective definition then requires operators to iterate over many possible objective weights and definitions, a process that can take many times longer than the optimization itself. A more versatile approach is multi-objective optimization, which establishes the trade-off curve or Pareto front between objectives. Here we present the first results on multi-objective Bayesian optimization of a simulated laser-plasma accelerator. We find that multi-objective optimization reaches comparable performance to its single-objective counterparts while allowing for instant evaluation of entirely new objectives. This dramatically reduces the time required to find appropriate objective definitions for new problems. Additionally, our multi-objective, multi-fidelity method reduces the time required for an optimization run by an order of magnitude. It does so by dynamically choosing simulation resolution and box size, requiring fewer slow and expensive simulations as it learns about the Pareto-optimal solutions from fast low-resolution runs. The techniques demonstrated in this paper can easily be translated into many different computational and experimental use cases beyond accelerator optimization.
\end{abstract}

\maketitle

\section{Introduction}

Laser-plasma interaction \cite{Macchi.2013, McKenna.2013} and in particular its sub-field of laser-plasma acceleration \cite{Esarey.2009, Wenz.2020} are highly researched areas with prospects for numerous scientific and societal applications \cite{Albert.2016, Bolton.2018}. Until the past decade, both experimental and numerical investigations in these fields were often based on single or a few laser shots and particle-in-cell simulations \cite{Malka.2012}, respectively. Since then, improvements in laser technology as well as computing hard- and software have made it possible to gather data for hundreds or thousands of different configurations in both experiments and simulations \cite{Goetzfried.2020,Kirchen.2021,Bohlen.2022}. This has sparked interest in using advanced techniques from computer science, particularly machine learning methods, which can deal more efficiently with large multi-dimensional data sets than human operators \cite{dopp2022data}.

Early examples include the use of genetic algorithms \cite{He.2015, Streeter.2018} and, more recently, the first measurements using surrogate models have been presented \cite{Shalloo.2020, Jalas.2021}. The latter are intermediate models that are generated based on existing data during optimization and that can be quickly explored numerically. Studies involving this \textit{Bayesian} optimization have demonstrated clear optimization of a carefully chosen optimization goal. Importantly, this goal has to be encoded in form of a so-called objective function, which acts on the measurement and gives a scalar output. In the case of a particle accelerator, the beam can generally be described by the charge distribution $\rho(\vec x,\vec p)$ in the six-dimensional phase space, and an objective function that optimizes beam parameters will act on this distribution or a subset of it. One of the simplest examples of an objective function is the charge objective function 
$$g_Q(\rho(\vec x,\vec p)) = \int \rho(\vec x,\vec p) d\vec x d\vec p.$$
While this function can \emph{in principle} be used as an objective function in a particle accelerator, it will usually not yield a useful optimization result. This is because it optimizes \textit{solely} the charge and all other beam parameters such as divergence and energy are lost in the integration process. In fact, due to energy conservation, this optimizer tends to reduce the beam energy, which is an unintended consequence in almost all conceivable applications of particle accelerators. 

In practice, one usually uses a combination of objectives, e.g. reaching a certain charge above a certain energy or the total beam energy. The design of objective functions for these problems is even more difficult because they need to give some constraints or limits to the single objectives. Many multi-objective scalarizations take the form of a weighted product $g = \prod g_i^{\alpha_i}$ or sum $g= \sum \alpha_i g_i$ of the individual objectives $g_i$ with the hyperparameter $\alpha_i$ describing its weight. For instance, \citet{Jalas.2021} optimized the spectrum of a laser-accelerated beam using an objective function that combines the beam charge $Q$, the median energy $\tilde E$ and the median absolute deviation $\Delta \tilde E$. Their proposed objective function to be maximized is $\sqrt{Q}\tilde E /\Delta\tilde{E}$, i.e. the exponential weights are $\alpha_1 = 0.5$, $\alpha_2 = 1$ and $\alpha_3 = -1$. Here, the use of median-based metrics will result in less sensitivity to outliers in the spectrum, while the weight parameter $\alpha_1 = 0.5$ explicitly reduces the relevance of charge compared to beam energy and spread. 

The choice of particular weights is, however, entirely empirical and usually the result of trial and error. An objective function is thus not necessarily aligned with the actual optimization goal, and one often needs to manually adjust the parameters of the objective function over multiple optimization runs. In essence, instead of scanning the input parameters of an experiment or simulation, the human operator will be scanning \textit{hyperparameters} of the objective function many times over. The less prior knowledge about the system is known, the longer this process may take.

The underlying problem is essentially one of compression, i.e. that the objective function needs to reduce a complex distribution function to a \emph{single number} characterizing said distribution. It is impossible to do this without information loss for an unknown distribution function. In fact, even if we knew the distribution, e.g. a normal distribution, one would still need \emph{both} mean and variance to describe it without ambiguities. In the case of an unknown one-dimensional distribution function, we can use multiple statistical descriptions to capture essential features of the distribution such as the central tendency (weighted arithmetic or truncated mean, the median, mode, percentiles, etc.) and the statistical dispersion of the distribution (full width at half maximum, median absolute deviation, standard deviation, maximum deviation, etc.). These measures weigh different features in the distribution differently. One may also include higher-order features such as the skewness, which occurs for instance as a sign of beam loading in energy spectra of laser-plasma accelerators \cite{Goetzfried.2020}, or coupling terms between the different parameters. Last, the amplitude or integral of the distribution function are often parameters of interest \footnote{One 'exception' may be applications that deal with probability distributions, which are by definition normalized.}. 

In the following, we will discuss optimizations of electron energy spectra according to different objective definitions and then present a more general multi-objective optimization. 

The paper is structured as follows: First, we are going to discuss details of the simulated laser-plasma accelerator used for our numerical experiments (\cref{System}) and introduce Bayesian optimization (\cref{sec:BO}). Then we present results from optimization runs using different definitions of scalarized objectives that aim for beams with high charge and low energy spread at a certain target energy (\cref{Single-Objective}). We then compare these results with an optimization using effective hypervolume optimization of all objectives (\cref{sec:Multi-Objective}). In \cref{Analysis} we discuss some of the physics that the optimizer 'discovers' during optimization and in the last section, we summarize our results and outline perspectives for future research (\cref{Summary}).

\section{Laser-plasma accelerator}\label{System}

\begin{table*}
\begin{tabular}{ ll|ll} 
\toprule
\multicolumn{4}{c}{\textbf{Variable input parameters}}\\
\midrule
\multicolumn{2}{c}{}& \textit{min. value} & \textit{max. value}\\
\textbf{Plateau Plasma density} & $n_e$ & $\SI{2e18}{\per\cubic\cm}$ & $\SI{9e18}{\per\cubic\cm}$\\
\textbf{Upramp length} & $l_{up}$ & $\SI{0.25}{\mm}$ & $\SI{1.75}{\mm}$ \\
\textbf{Downramp length} & $l_{down}$ & $\SI{0.0}{\micro\m}$ & $\SI{50}{\micro\m}$ \\
\textbf{Focus position} & $z_0$ & $\SI{-0.5}{\mm}$ & $\SI{2.5}{\mm}$ \\
\textbf{Simulation fidelity} & $\chi$ & 1 & 4 \\
\midrule
\multicolumn{4}{c}{\textbf{Fixed input parameters}}\\
\midrule
Laser wavelength & $\lambda_0$ & $\SI{800}{\nm}$ \\
Laser power & $P$ & $\SI{50}{\tera\watt}$ \\
Laser waist (FWHM) & $w_0^{FWHM}$ & $\SI{20}{\micro\m}$ \\
Laser duration (FWHM) & $\Delta t$ & $\SI{30}{\femto\s}$ \\
\midrule
\multicolumn{4}{c}{\textbf{Dependent variables}}\\
\midrule
Plasma wavelength & $\lambda_p$ & \multicolumn{2}{l}{$2\pi c \sqrt{m_e\epsilon_0/e^2n_e}$}\\
Plasma wavenumber & $k_p$ & $2\pi/\lambda_p$\\ 
Critical density & $n_c$ & \multicolumn{2}{l}{$(2\pi c/\lambda_0)^2(m_e \epsilon_0/e^2)$ }\\
Critical power & $P_c$ & \multicolumn{2}{l}{$2 m_e c^3 n_c/(r_e n_e)$ }\\
Peak intensity & $I_0$ &\multicolumn{2}{l}{ $2P/(\pi w_0^2)$} \\
Peak potential & $a_0$ & \multicolumn{2}{l}{$\sqrt{2I_0/\epsilon_0 c}\cdot 
(e/k_p m_e c^2)$} \\
Matched peak potential & $a_0^{matched}$ & \multicolumn{2}{l}{ $2(P/P_c)^{1/3}$}\\
Matched bubble radius & $r_b$ & \multicolumn{2}{l}{$\sqrt{2a_0^{matched}/k_p}$} \\
Rayleigh length & $z_R$ & $\pi w_0^2 / \lambda_0$\\
Waist  & $w$ & \multicolumn{2}{l}{$\sqrt{1+(z-z_0)/z_R)^2}$}\\
\multicolumn{2}{l}{(Gaussian beam in vacuum)} & & \\
\midrule
\multicolumn{4}{c}{\textbf{Simulation mesh parameters}}\\
\midrule
Transverse box size & $l_r$ & $2.5\cdot w(z=0)$\\
Longitudinal box size & $l_z$ & \multicolumn{2}{l}{$\SI{25}{\micro\m}+r_b$}  \\
Simulation length & $l_{z,max}$ & $\SI{3.5}{\mm}$ \\
Transverse resolution & $\Delta r$ & $\SI{600}{\nm}/\chi$\\
Longitudinal resolution & $\Delta z$ & $\SI{60}{\nm}/\chi$\\
Boost factor & $\gamma_{boost}$ & $\sqrt{l_{z,max}/l_z}/\chi$ \\ 
\bottomrule
\end{tabular}
\caption{\textbf{Simulation and scan parameters.} The top section shows the four simulation parameters and their ranges that are used in the optimization problem. Furthermore, a fidelity parameter $\chi$ is introduced that allows the optimizer to choose between low and high numerical resolution (see the section on mesh parameters). Based on the scan parameters and the fixed problem parameters, we calculate several dependent variables that help us to estimate the correct box size for the simulations.} 
\label{table1}
\end{table*}

\begin{figure}[t]
  \centering
  \includegraphics*[width=8cm]{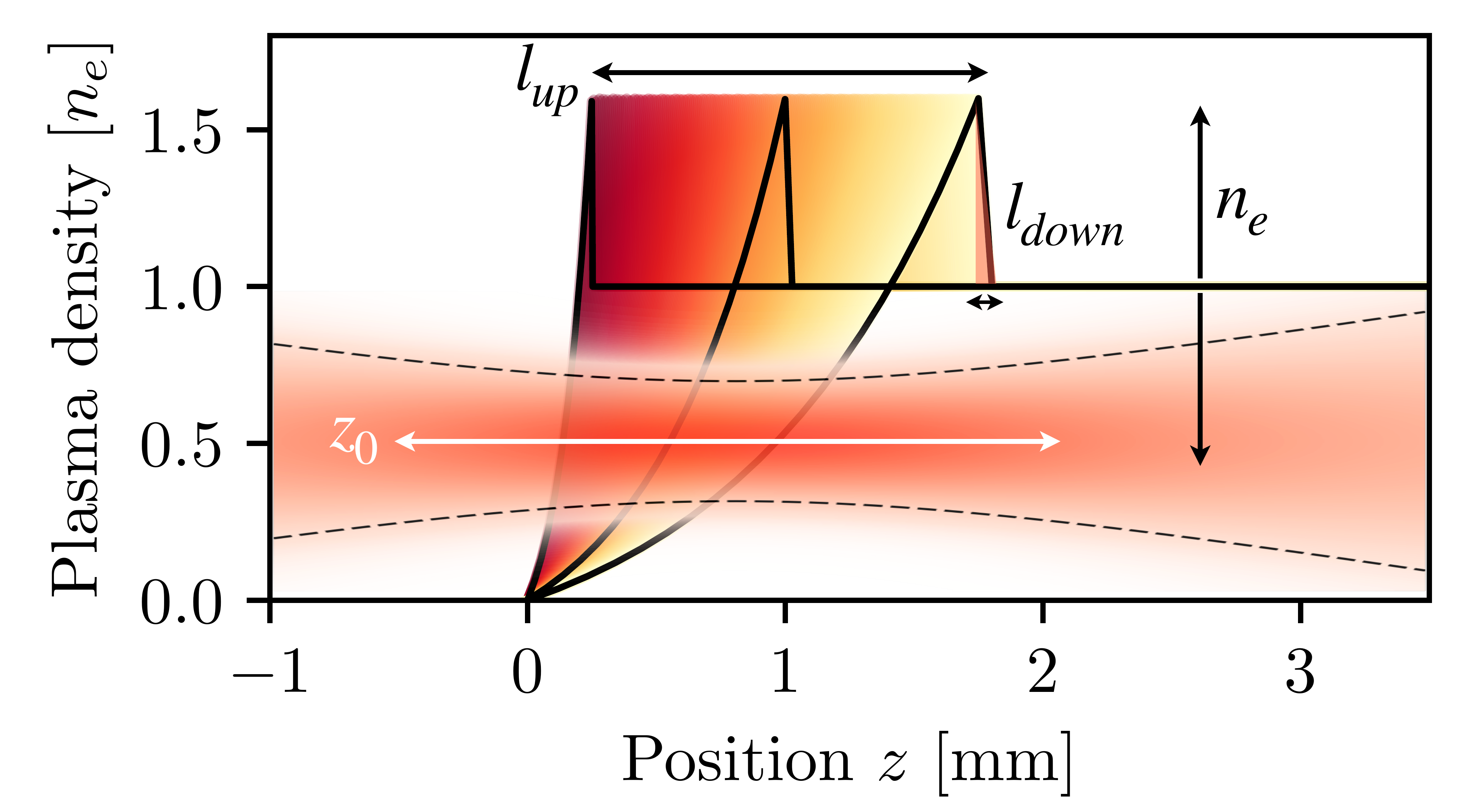}
  \caption{\textbf{Illustration of the four variable input parameters} from \cref{table1}, namely the upramp length $l_{up}$, the downramp length $l_{down}$, the plateau density $n_e$ and the focus position $z_0$.}
  \label{Input}
\end{figure}

As a test system for optimization, we use an example from the realm of plasma-based acceleration, i.e. a laser wakefield accelerator with electron injection in a sharp density downramp \cite{Buck.2013, Goetzfried.2020}. The basic scenario here is that electrons get trapped in a laser-driven plasma wave due to a local reduction in the plasma density, which is often realized experimentally as a transition from one side to the other of a hydrodynamic shock, hence the often-used name "shock injection". The number of electrons injected at this density transition strongly depends on the laser parameters at the moment of injection, but also on the plasma density itself. Both parameters also affect the final energy spectrum the electrons exhibit at the end of the acceleration process. Here we will use simulations to investigate this system, the primary reason being that they are perfectly reproducible and do not require additional handling of jitter, drifts, and noise. However, the methods outlined in this paper are equally relevant to experiments. The input space consists of four variable parameters, namely the plateau plasma density, the position of laser focus, as well as the lengths of the up- and downramps of the plasma density close to the density transition. 

While the shock injection scenario is sufficiently complex to require particle-in-cell codes, we use the code FBPIC by Lehe et al. \cite{Lehe.2016} in conjunction with various optimizations to achieve an hour-scale run-time. On the hardware side, the code is optimized to run on NVIDIA GPUs (here we used Tesla V100 or RTX3090), while the physical model includes optimizations such as the usage of a cylindrical geometry with Fourier decomposition in the angular direction and boosted-frame moving windows \cite{Kirchen.2016}. Additionally, we can take advantage of the very localized injection to locally increase the macro-particles density in the injection area \cite{Goetzfried.2020}. Similarly, the linear wakefields forming in regions of lower laser intensity result in a nearly laminar flow of particles, meaning that we can decrease the macro-particle density far away from the laser axis \cite{Ding.2020}.

One particular challenge that arises in simulations over a large range of parameters is that different input parameters may result in different computational requirements. For instance, the transverse box size needs to be several times larger than the beam waist to assure that the energy of a focusing beam is not lost. Hence, a laser that is initialized out of focus requires a larger box size than a beam initialized in focus. We address this by scaling the transverse box size $l_r$ as a function of the laser waist $w(z)$ at the beginning of the simulation. Similarly, the size of the wakefield depends on the plasma density, and accordingly, we scale the longitudinal size $l_z$ of the box with the estimated wakefield size. By using these adapted simulation boxes, we avoid wasting computational resources and only capture the physics relevant to our problem. It should be noted that the scan range of these highly optimized simulations is to some extent limited by the appearance of numerical instabilities or artifacts and, for instance, the boosted-frame geometry cannot be used in a near-critical setting. A summary of all free and dependent parameters of the simulations is given in \cref{table1} and illustrated in \cref{Input}.

\section{Bayesian optimization}\label{sec:BO}

In this paper, we make use of Bayesian optimization, a sequential model-based algorithm, to optimize problems that are either costly or time-intensive to probe. Bayesian optimization works by constructing a probabilistic surrogate model of an objective function by sampling the parameter space, see Döpp \emph{et al.}\cite{dopp2022data} for an overview. In this paper, we use a Gaussian process (GP) as the probabilistic model, which is a non-parametric model based on a prior (in this case a zero-mean distribution) and a covariance function that expresses the correlation between the prior and the current data points. This surrogate model is cheap and fast to evaluate, and Bayesian optimization finds the next evaluation point by optimizing the model instead of the real system. To this end, a so-called acquisition function, which quantifies the expected improvement from a certain set of input parameters, is used. After a measurement at this point, the model is updated, and the procedure is repeated until a certain stopping criterion is fulfilled.

The main advantage of Bayesian optimization compared to other methods is that it can find the global optimum of a function in a very sample-efficient way. Furthermore, the acquisition function and the model can be adapted to not only optimize a single objective but multiple combinations of objectives. This adaptation is done in multi-objective optimization (see \cref{sec:Multi-Objective}), where Bayesian optimization can implicitly optimize multiple combinations of objectives by optimizing the \emph{expected hypervolume improvement}\cite{yang2019multi,daulton2020differentiable}.

Despite its high sample efficiency, optimizing a multi-dimensional problem still requires a non-negligible number of evaluations. For our laser-plasma accelerator with four variable inputs we typically need to perform on the order of $\sim 10^2$ evaluations to locate the optimum. Given the hour-scale runtime of our simulations discussed in \cref{System}, a full optimization run would that take several days to compute. 

We can speed up the optimization process and allow for multi-dimensional optimization by using low-resolution simulations that use a larger numerical grid, see \cref{table1}, and a larger boost factor. They capture the essential physics of injection and acceleration but have not yet fully converged in terms of final charge, energy, and so forth. These approximate solutions take only a few GPU minutes to compute while providing valuable information for optimization. Importantly, we can directly incorporate the possibility of varying the resolution and hence, \emph{fidelity} of a simulation into the optimization process by introducing a new fidelity variable $\chi$ (see \cref{table1}). In a process called multi-fidelity optimization, we construct a Gaussian process that models the objective function over the four input dimensions ($n_e$, $l_{up}$, $l_{down}$, $z_0$) as well as the fidelity parameter $\chi$. The decision regarding the next position to probe is taken by a recently introduced multi-objective, multi-fidelity (MOMF) Bayesian optimization algorithm \cite{Irshad.2021}, which is based on the common optimization of the different objectives and an additional trust objective. Regarding the latter, the algorithm also considers the computational cost associated with the fidelity parameter. From a convergence study of PIC simulations, we found that the computing time of our simulations approximately scales with $cost(\chi)\propto\chi^{3.5}$. The speed-up gained by taking this cost and fidelity information into account is on average an order of magnitude in this study and a full multi-fidelity optimization run typically takes about 10 hours. 

While designed for multiple objectives, the MOMF acquisition function also supports the optimization of a single objective with multiple fidelities \cite{Irshad.2021}. All results presented in the following, both single- and multi-objective, are thus obtained using the same algorithm for a fair comparison. The optimization parameters are outlined in \cref{table3} and importantly, both objective types have the same constraints regarding maximum iteration number and computational budget, as well, and they were executed 5 times with 5 random initial points each to assess the typical performance. 

\begin{table}[t]
\begin{tabular}{ ll|l} 
\toprule
\multicolumn{3}{c}{\textbf{Optimization parameters}}\\
\midrule
Number of Trials &  $n_{TRIALS}$ & 5\\
Max. number of Iterations & $n_{BATCH}$ & 150\\
Maximum Cost &  $C_{total}$ & 50 GPU hours\\
Number of initial points & $n_{INIT}$ & 5\\
Input Dimensions & $dim_x$ & 5\\
Output Dimensions & $dim_y$ & 1 (single objective) or \\
& & 3 (multi-objective)\\
Cost Function & $cost(\chi) $ & $\propto\chi^{3.5}$\\
\bottomrule
\end{tabular}
\caption{\textbf{Summary of optimization parameters used in the paper} Some of the parameters used to run the MOMF algorithm are outlined in this table. The number of iterations and total cost are two upper thresholds used to stop the optimization run. When either value was reached the optimization was stopped. The output dimensions for scalarized runs was 1 while the MOMF optimized 3 objectives simultaneously. The cost function is approximated due to the adaptive meshes used in this study. } 
\label{table3}
\end{table}

\begin{table*}[t!]
\begin{tabular}{ ll|l} 
\toprule
\multicolumn{3}{c}{\textbf{Objective definitions}}\\
\midrule
Objective 1 &  $O_1$ & ${Q^{0.5}}((\Delta \bar E^{2}) \sigma_E)^{-1}$, \cref{Obj1}\\
Objective 2 &  $O_2$ & ${Q^{0.5}}((|\Delta \tilde E|) E_{MAD})^{-1}$, \cref{Obj2}\\
Objective $2_{a}$ &  $O_{2a}$ & ${Q^{2}}((|\Delta \tilde E|) E_{MAD})^{-1}$, \cref{Obj21}\\
Objective $2_{b}$ &  $O_{2b}$ & ${Q^{3}}((|\Delta \tilde E|) E_{MAD})^{-1}$, \cref{Obj22}\\
Objective 3 &  $O_3$ & $2{Q_{in}}-{Q}$, \cref{Obj3}\\
\midrule
\multicolumn{3}{c}{\textbf{Charge-related metrics}}\\
\midrule
$Q$ & \multicolumn{2}{l}{Total integrated charge} \\
$Q_{in}$ & \multicolumn{2}{l}{Charge within an energy interval $E_0\pm \Delta E$} \\
\midrule
\multicolumn{3}{c}{\textbf{Central tendency metrics}}\\
\midrule
$\bar E$ & \multicolumn{2}{l}{Mean energy }\\
$\tilde E$ &\multicolumn{2}{l}{Median energy }\\
$E_0$ & \multicolumn{2}{l}{Target energy (300 MeV)}\\
$\Delta \bar E^{2}$ & \multicolumn{2}{l}{Mean-squared difference of median and target energy}\\
$|\Delta \tilde E|$ & \multicolumn{2}{l}{Absolute difference of median and target energy}\\
\midrule
\multicolumn{3}{c}{\textbf{Statistical dispersion metrics}}\\
\midrule
$\sigma_E$ & \multicolumn{2}{l}{standard deviation }\\
$E_{MAD}$ & \multicolumn{2}{l}{median absolute deviation }\\
\bottomrule
\end{tabular}
\caption{\textbf{Summary of single-objective functions used in the paper.} The five single-objective scalarized functions that are optimized in this study are shown at the top. The middle and bottom parts show the charge, central tendency, and statistical dispersion metrics used to construct the single objectives. These are also used in multi-objective multi-fidelity optimization. } 
\label{table2}
\end{table*}

\section{Single-objective optimization}\label{Single-Objective}

The goal of the optimization presented here is to produce quasi-monoenergetic electron beams with a high total charge around a certain target energy $E_0$. In statistical terms, these goals can be captured by the difference of the central tendency from the target energy, statistical dispersion, and the integral of the electron beam spectrum. But as mentioned in the introduction, these three features can be described by multiple statistical measures such as the standard deviation, median absolute deviation, mean energy, median energy, and total charge \cite{Ruppert.2011}. In practice there thus exists large freedom how exactly these objectives are encoded into a single \emph{scalarized} objective. Each objective function has a bias toward a particular outcome and thus, the final optimization result may differ significantly. In the following, we present several different objective functions that intend to reach the same goal, i.e. simultaneously maximize charge, reduce spectral width and reduce the distance to the target energy.

\textit{Examples.} In terms of the mean energy and standard deviation we can define the objective
\begin{equation}
    O_1=\frac{Q^{\frac12}}{\Delta \bar E^{2} \cdot \sigma_E}
    \label{Obj1}
\end{equation}
where $Q$ is the total charge, $\Delta \bar E^2=|\bar E - E_{0}|^2 + \epsilon$ is the squared difference between the mean energy $\bar E$ of the spectrum and the target energy and $\sigma_E$ is the standard deviation. Note that $\epsilon$ in the definition of $\Delta \bar E^2$ is an offset to prevent the objective from approaching infinity as the distance to the target energy is decreased. Throughout the manuscript we use $\epsilon=\SI{1}{\MeV}$, as beams within a distance of $\SI{1}{\MeV}$ to the target energy are considered sufficiently optimized.

It is a characteristic of the mean that it tends to emphasize points further away from the target. In presence of noise, it is thus often suitable to use median-based descriptors, instead. Such an objective could be
\begin{equation}
    O_2=\frac{Q^{\frac12}}{|\Delta \tilde E|\cdot  E_{MAD}},
    \label{Obj2}
\end{equation}

where $Q$ is the total charge, $\tilde E$ is the median energy of the spectrum, $E_{0}$ is the target energy, $|\Delta \tilde E| = |\tilde E-E_0|+\epsilon$ is their absolute distance (plus offset) and $E_{MAD}$ is the median absolute deviation around the median. Note the use of the square root to decrease the emphasis placed on the total charge. This is essentially the aforementioned objective used by Jalas \emph{et al.}\cite{Jalas.2021}, with the difference that we use a target energy instead of an energy maximization.

The choice of $Q^{1/2}$ is, however, entirely empirical and we can equally well define alternative versions of such an objective function with different exponential weights of charge. For instance, we can use the two objectives

\begin{equation}
    O_{2,a}=\frac{Q^{2}}{|\Delta \tilde E|\cdot E_{MAD}},
    \label{Obj21}
\end{equation}
and
\begin{equation}
    O_{2,b}=\frac{Q^{3}}{|\Delta \tilde E|\cdot E_{MAD}},
    \label{Obj22}
\end{equation}
which should incentivise the optimizer to find beams with higher total charge.

We already alluded to the general problem that objectives using division can get arbitrarily large when decreasing the value of the denominator. Instead of circumventing this problem with offsets, it can be preferable to rewrite the objective entirely without using division. One possible way to do so in our case is to implicitly optimize the target energy and energy spread by optimizing the charge within a certain energy window. This can be written in the form 
\begin{equation}
    O_3= 2{Q_{in}}-{Q}, 
    \label{Obj3}
\end{equation}
where $Q_{in}=\int_{E_0-\Delta E/2}^{E_0+\Delta E/2} Q(E) dE$ is the charge within a given energy interval $\Delta E$ around the target energy. A summary of these single objectives and definitions of metrics used to define them is outlined in \cref{table2}.

\begin{figure}[t]
  \centering
  \includegraphics*[width=8.5cm]{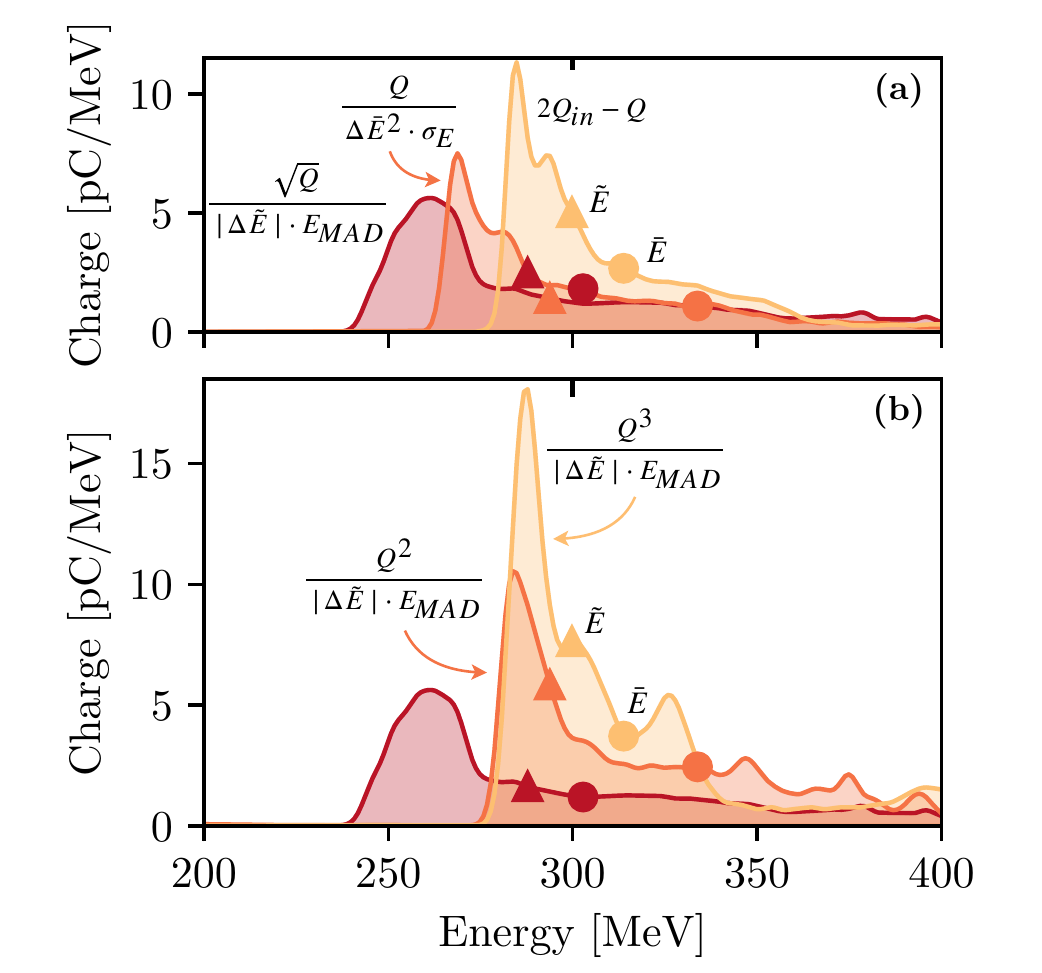}

  \caption{\textbf{Single-objective spectra.} (a) Final spectra obtained using three different objectives ($O_1$, $O_2$ and $O_3$ from the text) to optimize beam charge, beam distance from target energy ($\SI{300}{\MeV}$) and energy spread. Median energy $\tilde E$ of each spectrum is indicated using triangular markers, while mean energy $\bar E$ is marked with circles. (b) Example for changes in objective weight. Here we use variations of the $O_2$ objective with charge squared ($O_{2,a}$) or charge to the power of three ($O_{2,b}$), leading to higher overall charge in the beam and - without explicit optimization - more peaked spectra.}
  \label{Spec_single}
\end{figure}

\textit{Results and discussion.} Having defined several "sensible" objectives, we now present results using them to optimize the simulated laser wakefield accelerator.
In \cref{Spec_single}, we show the final spectrum of the three scalarized objectives $O_1$, $O_2$ and $O_3$. Since we use mean energy for the first objective the spectrum outliers can influence the mean much more than the second objective using median energy. 
This explains why the first spectrum tends to not have any high or low-energy tail. The second spectrum using median energy and median absolute deviation allows having a high energy tail while keeping the median close to 300 MeV. Since this spectrum has a longer tail the mean of this spectrum is higher than 300 MeV. The third objective in \cref{Spec_single} has a much higher peak charge because it has a higher implicit weighting on charge. The beam here has an even shorter tail since it explicitly penalizes charge outside of the $250-350$ MeV window. Overall, we can see that the different scalarization of the statistical measures can result in different spectra. Also, implicitly optimizing for beams near a target energy, as is done in the case of the third objective, appears to yield better results than explicitly optimizing for it. 

One prominent feature in these spectra is that the energy exhibiting the highest spectral charge density of electrons, subsequently referred to as peak energy $E_{peak}$, is much further from the target energy than the mean or median energies. This is because highly-charged electron beams create beam loading effects in laser wakefield accelerators, resulting in skewed spectra \cite{Goetzfried.2020}. For such asymmetric spectra, the peak does not coincide with the distribution's mean or median and an explicit optimization of the peak energy becomes necessary, which we discuss in more detail in the next section.

In a next step we compare the different versions of the second objective ($O_2$, $O_{2,a}$ and $O_{2,b}$) with $Q^{1/2}$,$Q^{2}$ and $Q^{3}$ weighting, respectively. As expected, the higher weight increases the total charge in the optimized beam spectrum. 
However, while this particular combination of hyperparameters appears to yield objectively better beams than the $O_{2}$ and $O_{2a}$ variations, it is not possible to know this beforehand, adding the hyperparameter choice as an additional degree of freedom to the optimization problem. For any new problem an operator or user thus needs to run several optimization run to identify the most suitable definitions and combinations of objectives.


\begin{figure*}[t]
  \centering
  \includegraphics*[width=.9\linewidth]{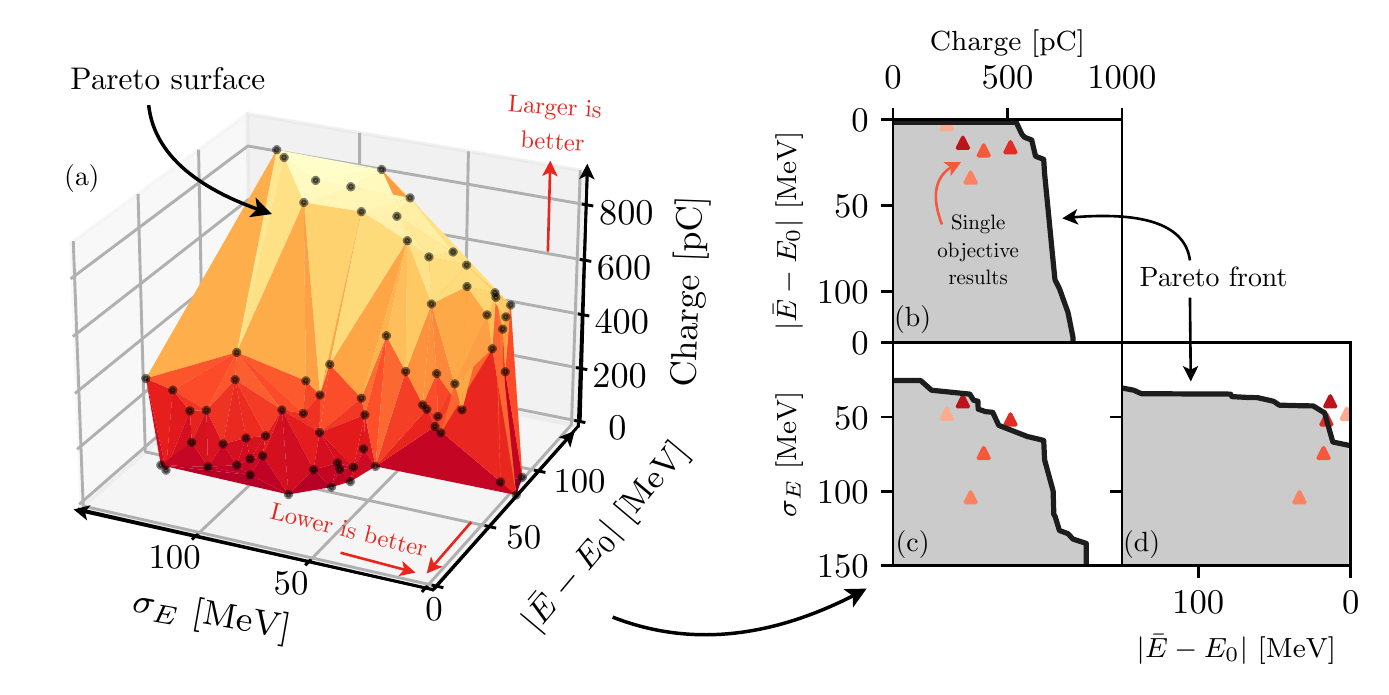}

  \caption{\textbf{Multi-objective optimization.} (a) Visualization of the Pareto surface spanned by  the non-dominated solutions for each of the three objectives. (b-d) 2-D projections of the Pareto surface, showing the Pareto front for the objective pairs of charge vs. energy distance (b), charge vs. energy spread (c) and energy spread vs. energy distance (d). The results show that the results of a single multi-objective optimization are either similar or better than all of the single-objective runs.}
  \label{fig:pareto-surface}
\end{figure*}


\begin{figure}[tb]
  \centering
  \includegraphics*[width=8cm]{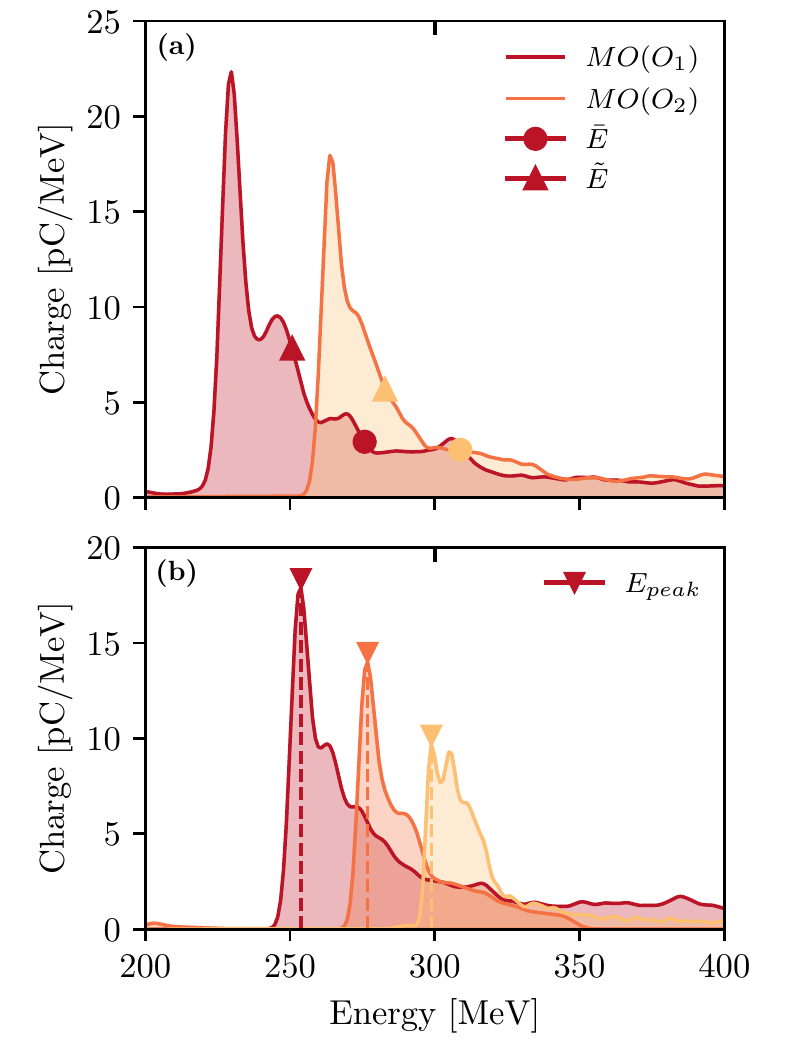}

  \caption{\textbf{Selected spectra obtained via one multi-objective optimization run.} (a) Spectra selected as optimal lower confidence bound solutions for the objectives $O_1$ and $O_2$. (b) Solutions optimized for peak energies of \SI{250}{\MeV}, \SI{275}{\MeV} and \SI{300}{\MeV}. }
  \label{fig:MO_solutions}
\end{figure}

\section{Multi-objective optimization}\label{sec:Multi-Objective}

As we have seen in the previous section, a major problem with the single-objective optimizations of complex systems is that the optimal weights for the hyperparameters in single-objective optimization are not known \emph{a priori}. Thus, to get a higher value for one particular objective, the weights need to be changed via trial and error. Furthermore, multi-objective optimization problems often exhibit some trade-offs in the optimization of different objectives. As a result, changing one objective's weight will also affect the other objectives, in either a detrimental or beneficial way. A single-objective optimization will always be biased towards a particular trade-off. But it is difficult or impossible to assess this bias beforehand, and the optimization will often not yield the optimal trade-off of parameters a user or operator intended. 

A more versatile strategy is to directly explore the trade-off between different objectives and choose the optimal combination of objectives \emph{a posteriori}. This trade-off optimization results in a solution set that in the output space is known as the \emph{Pareto front} and in the input space as the \emph{Pareto set}. A point is said to dominate another when it has at a minimum higher value for one objective keeping others equal. Thus, the Pareto front is the set of non-dominated points in any given output space. The area covered by the dominated space is known as the hypervolume and it is an indirect measure for the diversity of solutions. In Bayesian optimization the expected hypervolume improvement can therefore be used to optimize different objectives simultaneously. In our case we choose the mean energy difference $\Delta\bar E = |\bar E-E_{0}|$, the standard deviation $\sigma _E$ and total charge in the beam $Q$ as individual objectives spanning the output space. 

\textit{Results and discussion.} In \cref{fig:pareto-surface} we show the results for one representative run of the multi-objective Bayesian optimization. By probing the Gaussian process model, we obtain an entire set of solutions that can be visualized as a Pareto surface, consisting of all the non-dominated points in the three-dimensional output space. Projections showing the Pareto fronts for the three pairs of objectives are shown in \cref{fig:pareto-surface}b-d. We also indicate the beam parameters of the different optimizations presented in the previous section as blue triangles. The results show that the multi-objective optimization yields comparable performance to the combinations of objectives discussed in \cref{Single-Objective}.

This figure also shows some trade-offs, inherent to many multi-objective problems, some of which have an underlying physics interpretation. One prominent feature can be seen in \cref{fig:pareto-surface}b, where an increase in the distance to target energy is seen when the total charge exceeds $\SI{500}{\pico\coulomb}$. This distance is mainly due to a decrease in energy caused by beam loading \cite{Goetzfried.2020}: As the charge increases the electron bunch dampens the strength of the wakefields, which consequently leads to lower mean energy and thus, an increase in the distance to the target energy. Another trade-off in \cref{fig:pareto-surface}c is between high charge and mono-energetic beams, where increasing charge results in a wider spectrum of the electron beams. This indicates that the input parameters that yield a beam with a higher total charge are different from the ones that produce quasi-mono-energetic beams, an effect seen throughout the literature, e.g. in Götzfried et al.\cite{Goetzfried.2020}. Another notable result, albeit not directly visible from the plots, is the absence of any high-energy beams with low charge. This is because high-energy beams are implicitly excluded by the three objectives. As the energy increases, the distance to the target energy increased and the charge injected for these high-energy beams is less than those for lower-energy beams. Hence, most beams are restricted to energies near or lower than the target energy.

The effect of exploring inherent trade-offs makes multi-objective Bayesian optimization a very useful optimization technique. Two important benefits are (a) that one can characterize the capability of the system (laser-plasma accelerator) regarding each objective and (b) that it yields a solution set without strong bias towards particular objective combinations. The expression of the objective in terms of the hypervolume also avoids the problem seen in single-objective optimization, where offset values needed to be included for objectives in the denominator. This is because the hypervolume does not increase drastically if only a single objective is concerned and hence, multi-objective optimization does not excessively exploit single objectives.

As mentioned earlier, another important feature of this optimization strategy is that the Gaussian process model is cheap to probe and provides immediate feedback regarding the predicted means and variances of each individual objective (here $Q$, $\sigma_E$ and $\Delta\bar E$) for a combination of input parameters $x$ (i.e. $n_e$, $l_{down}$, $l_{up}$ and $z_0$). These can easily be combined into any desired objective $O(x)$. We can propagate the variances through the new objective to get an estimate of the uncertainty\footnote{For a generic objective of the form $O(x)=x_1/(x_2\cdot x_3)$ the uncertainties $\sigma({x_i})$ propagate as $\frac{\sigma(O(x))}{\mu(O(x))}\approx \sqrt{\left({\frac  {\sigma(x_1)}{\mu(x_1)}}\right)^{2}+\left({\frac{\sigma({x_2})}{\mu({x_2})}}\right)^{2}+\left({\frac{\sigma({x_3})}{\mu({x_3})}}\right)^{2}}$
} and a conservative solution candidate $\hat x$ can be found using the lower confidence bound 
\begin{equation}
\hat x =   \underset{x}{\mbox{argmax}} \{\mu(O(x)) - \sigma (O(x))\}.
\end{equation}

In \cref{fig:MO_solutions}a we show such inferred solutions for the previously defined objectives $O_1$ and $O_2$, see \cref{Obj1} and \cref{Obj2}, respectively. Due to the higher charge in these beams, the value of $O_1$ is approximately 40\% higher than in the single-objective optimization result (see \cref{Spec_single}a). The results for $O_2$ are comparable to those shown before, with the multi-objective result reaching 90\% of the corresponding single-objective result. The objective value is nonetheless diminished, most likely because the optimizer is not strictly optimizing the median energy but the mean energy.

It is notable that the spectral peaks in these candidate solutions are located at \SI{230}{\MeV} and \SI{272}{\MeV}, respectively, and thus, far from the "target" energy $E_0 =\SI{300}{\MeV}$. As discussed before, this happens because the mean or median of these highly skewed spectral distributions does not coincide with peak energy $E_{peak}$. We can address this problem without needing to run a new optimization. Instead, we take the existing multi-objective scan and construct a Gaussian process that predicts the peak energy for a given input $x$. Next, we can select suitable candidates using 
\begin{equation}
\hat x =   \underset{x}{\mbox{argmin}} \{\|E_{0}-\mu(E_{peak}(x))\| + \sigma (E_{peak}(x))\}.
\end{equation}
As this is a minimization problem, we now use the lower confidence bound to exploit the existing solutions while taking into account uncertainty. Suitable results are either found immediately, or after $1-2$ iterations for which the results of a candidate solution are used to improve the Gaussian process regression. The results of this process are shown in \cref{fig:MO_solutions}b, showing that the multi-objective results can even translate to objectives that differ substantially from the three objectives directing the hypervolume search. The results thus show that multi-objective Bayesian optimization greatly facilitates the process of finding both optimal parameter settings and optimal objective configurations. The latter can be evaluated \textit{a posteriori} at negligible cost and feed subsequent iterations of single-objective optimizers focused on objective exploitation. It should be emphasized that this process is most suitable in the optimization of unknown systems, e.g. newly set up experiments or simulations. For well-known systems the optimization of a well-working objective function, e.g. \cref{Obj3} with an appropriately chosen energy window, can produce competitive results.

\section{Input Space Analysis}\label{Analysis}
\begin{figure}[thp]
  \centering
  \includegraphics[width=.9\linewidth]{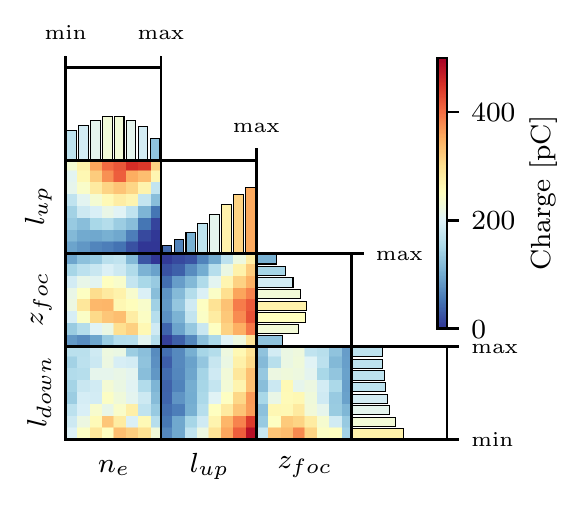}
  \includegraphics[width=.9\linewidth]{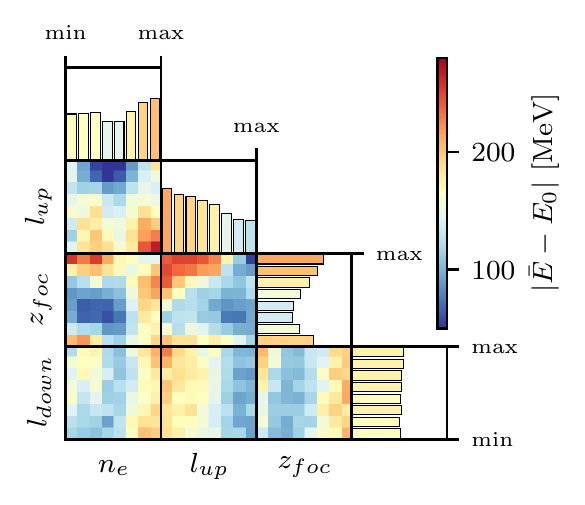}
  \includegraphics[width=.9\linewidth]{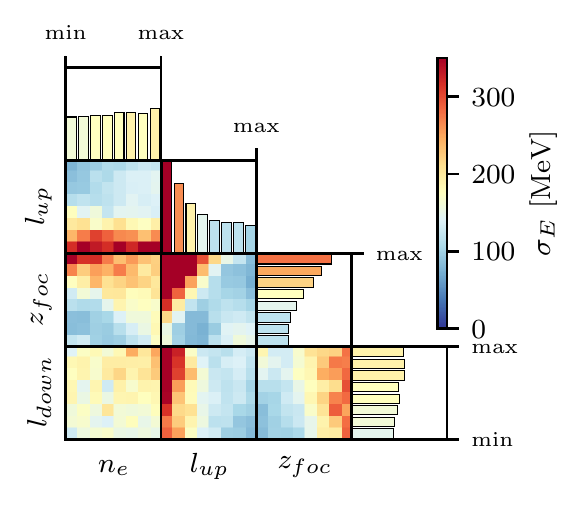}
  \caption{\textbf{Input space visualization.} The pair plots show the relationship between the four input parameters ($n_e$,$l_{down}$,$l_{up}$ and $z_0$) regarding the three outputs (charge, energy distance and energy spread). The 1D histograms show the average effect of each input parameter averaged.}
  \label{fig:input_space}
\end{figure}

So far, our discussion has been limited to the performance regarding single or multiple objectives. Another advantage of Bayesian optimization is that the model created during optimization can itself be analyzed and provide insights about the underlying physics and parameter dependencies of the system. In this section, we will thus look how specifically the choice of input parameters affects each of the different objectives. To do so, we use the data from our multi-objective optimization runs and train a GP model. 
We can sample points from this model to generate plots that map the influence of the input parameters on the individual objectives, either in a pair-wise comparison or individually. The results are shown in \cref{fig:input_space}. The data is binned by taking pair-wise input parameters while averaging over the other two. This results in a total of six two-dimensional plots for the four input parameters, which are color-coded by the values of the output objective. We can also see the influence of a single parameter by repeating the same procedure as above by averaging over the three input parameters. This is repeated three times to generate plots for each output objective.  Based on these plots we can observe trends for each input parameter, which in some cases have straightforward physical interpretations. 

\textit{Density ($n_e$)}: We observe that the charge initially increases with density, but then somewhat surprisingly starts to decrease again. Closer analysis of the underlying PIC simulations showed that injection indeed increases monotonically within our density range, but at higher density parts of the injected electrons are lost at the end of the accelerator due to dephasing and defocusing fields. We observe that the optimal energy is reached at the same density as the optimal charge, which indicates an optimal charge for beam loading to reach the desired target energy of \SI{300}{\MeV}. The energy spread tends to increase with the density and shows an interesting correlation with the focal plane.

\textit{Upramp length ($l_{up}$)}: We observe that the injected charge increases linearly with the upramp length. This effect is most likely related to laser self-focusing, where a longer upramp allows for stronger self-focusing of the laser and hence, a higher laser intensity at the injection point. We also observe that longer upramps seem to facilitate reaching the target energy.

\textit{Downramp length ($l_{down}$)}: The length of the downramp directly affects the injected charge, which is expected as shorter downramps corresponding to a more rapid wakefield expansion and thus increased injection. Meanwhile, the downramp length shows no influence on the mean energy, which also is to be expected because within the ranges scanned here it controls the point of the injection and not the acceleration length. 

\textit{Focus position ($z_{foc}$)}: The charge is maximized for a specific focus position, but for the same position we tend to observe the worst energy and energy spread, most likely because higher charge leads to beam loading.

For some parameter combinations we also observe couplings. For instance, the optimal focus position for maximal charge moves to the front the higher the plasma density and the longer upramp length. \\

\section{Conclusion and outlook}\label{Summary}

To conclude, we have presented the first multi-objective optimization of a simulated laser-plasma accelerator. The performance of the multi-objective optimizer was benchmarked against several single-objective optimizers and it was found to lead to similar or even superior results. Meanwhile, the multi-objective optimizer yields a far more general result that does not require iterative fine-tuning of objective parameters. 

By combining a state-of-the-art GPU-based simulation code with a multi-fidelity optimization algorithm, we were able to perform extensive, multi-dimensional optimizations that are to our knowledge without precedent in the field of notoriously expensive particle-in-cell simulations of physical systems. This result is a milestone towards using 'digital twins' of complex physical systems to optimize real-life experiments and infrastructure. This does not only concern laser-plasma acceleration but applies to any optimization problem with different available numerical resolutions.

While the multi-fidelity component most immediately benefits simulation studies with different resolutions, the multi-objective techniques presented in this paper can also be directly transferred to experiments. As already noted by \cite{Shalloo.2020} and discussed in \cref{Analysis}, surrogate models carry significant information about the physics interaction that can help physicists to better understand couplings between input and output parameters. This information may directly be used to improve experiments, e.g. as a prior for Bayesian optimization in experiments.
\begin{acknowledgments}
This work was supported by the DFG through the Cluster of Excellence Munich-Centre for Advanced Photonics (MAP EXC 158), TR-18 funding schemes and the Max Planck Society. It was also supported by the Independent Junior Research Group "Characterization and control of high-intensity laser pulses for particle acceleration", DFG Project No.~453619281. F.I. is part of the Max Planck School of Photonics supported by BMBF, Max Planck Society and Fraunhofer Society.
\end{acknowledgments}

\bibliography{ref}

\end{document}